\documentclass{article}
\usepackage{spconf,amsmath,graphicx,url}
\usepackage[font=small]{caption}
\usepackage{booktabs}
\usepackage{soul}
\usepackage[breaklinks]{hyperref}

\title{Small footprint Text-Independent Speaker Verification \\ for Embedded Systems}
\name{Julien Balian, Raffaele Tavarone,  Mathieu Poumeyrol, Alice Coucke}

\address{
\{firstname.lastname\}@sonos.com\\
Sonos Inc., Paris, France
}

\begin{document}

\maketitle
\begin{abstract}
Deep neural network approaches to speaker verification have proven successful, but typical computational requirements of State-Of-The-Art (SOTA) systems make them unsuited for embedded applications. In this work, we present a two-stage model architecture orders of magnitude smaller than common solutions ($237.5$K learning parameters, $11.5$MFLOPS) reaching a competitive result of $3.31$\% Equal Error Rate (EER) on the well established VoxCeleb1 verification test set. We demonstrate the possibility of running our solution on small devices typical of IoT systems such as the Raspberry Pi 3B with a latency smaller than $200$ms on a $5$s long utterance. Additionally, we evaluate our model on the acoustically challenging VOiCES corpus. We report a limited increase in EER of $2.6$ percentage points with respect to the best scoring model of the 2019 VOiCES from a Distance Challenge, against a reduction of $25.6$ times in the number of learning parameters.
\end{abstract}

\section{Introduction}
\label{sec:intro}
Speaker verification refers to the task of verifying a user identity based on their voiceprint. 
This technology has received increasing attention in recent years, partially due to its application to voice assistants. Speaker verification enables the contextualisation of spoken queries and tailored assistant responses to personalised content (\textit{e.g.} ``add an event to \textit{my} calendar``).

The aggregation of the variable-length sequential audio input into a fixed length embedding plays a crucial role for any practical application. In the first approaches to time aggregation, the embeddings are computed over fixed segments of audio~\cite{ivector} with additional steps required to aggregate the representations. More recently, with the advent of neural networks, end-to-end solutions have been proposed that directly handle variable-length input~\cite{heigold_end--end_2015}. Using properly designed layers, the temporal statistics can be accumulated internally in the network~\cite{heigold_end--end_2015,sp,asp}.
While reaching good accuracy, these end-to-end, neural-based methods have typical computational footprints that require offline or server-side execution. 
Although some speaker verification engines with low execution latency~\cite{ge2e}  or the ability to run on mobile devices~\cite{angular_prototypical}  have been proposed, 
they remain too large for embedded applications where memory and computing power are further limited.

In this work, we propose a speaker verification system specifically tailored to embedded use cases.
We budget CPU and memory resources to match that of typical keyword spotting systems~\cite{snips_wakeword} designed to run continuously and in real time on device. 
Our approach allows to decouple streamed time-series features extraction from aggregation, providing an optimal balance between representation quality and inference latency. 
The features extraction stage is based on the QuartzNet~\cite{quartznet} model -- never used in the context of speaker verification to our knowledge -- and the aggregation stage on Ghost Vector of Locally Aggregated Descriptors (GVLAD)~\cite{gvlad}, with key modifications, \textit{e.g.} the inclusion of Max  Features Map~\cite{mfm} operations and a more computationally efficient method for descriptor aggregation.

We demonstrate that our approach reaches performances comparable with the state of the art (3.31\% EER on the VoxCeleb1 verification test set) with a number of learning parameters orders of magnitude smaller, making it fit for embedded applications. Voice being a highly sensitive biometric identifier, our lightweight approach grants speaker verification abilities to small devices typical of IoT systems, while fully respecting the privacy of users.

The paper is structured as follows.  In Section ~\ref{sec:spkidmodel}, we detail our system in terms of neural network architectures and training method.
In Section~\ref{sec:experiments}, we describe our experimental settings including datasets and hyper-parameters selection, and computational performances. In Section~\ref{sec:results}, we compare our solution with other SOTA approaches. We give our conclusions in Section~\ref{sec:conclusion}.

\begin{figure*}[ht!]
    \centering
    \includegraphics[height=8.9cm]{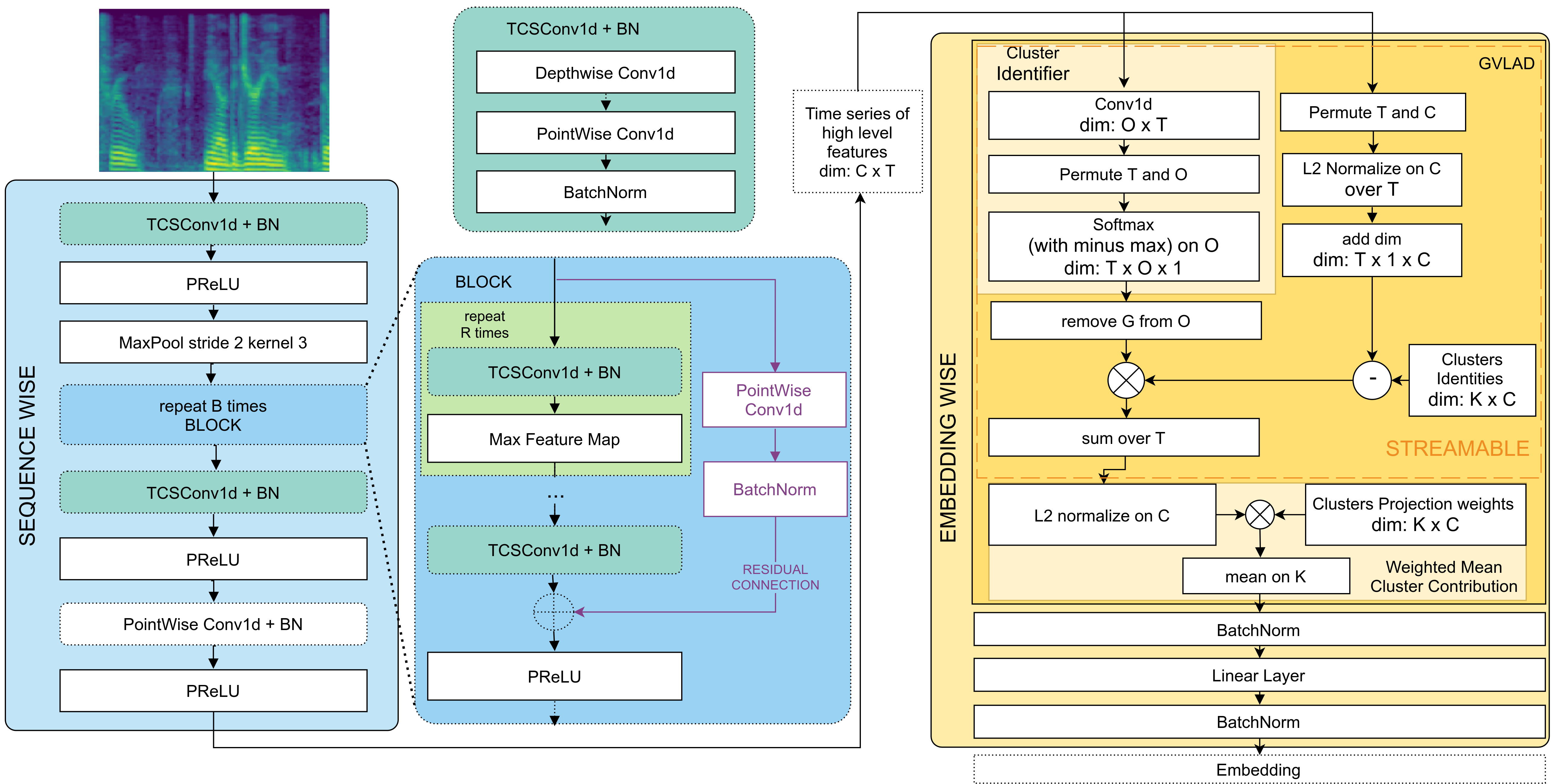}
    \caption{Network architecture. Left panel: sequence-wise network (blue). $B$ is the number of stacked blocks. In each block, $R$ is the number of TCSConv1d+BN modules followed by Max Feature Map activation stacked before a last TCSConv1d+BN and residual connection. Right panel: embedding-wise network (yellow). $T$ is the time dimension and $C$ is the channel dimension. In the GVLAD module, $O=K+G$ is the total number of clusters, defined as the sum of the number of contributing clusters $K$ and the number of Ghost clusters $G$.}
    \label{fig:network_diagram}
\end{figure*}

\section{Speaker Embedding Model}
\label{sec:spkidmodel}

We propose a fixed-size speaker embedding model performed in two stages. In the first stage, a streaming neural network, inspired by the QuartzNet~\cite{quartznet} architecture, takes as input an arbitrarily long time series of acoustic features and outputs another series of higher-level features. This network is referred to as sequence-wise network. The second stage consists of an aggregator neural network, called embedding-wise network, built upon the GVLAD~\cite{quartznet} architecture, that aggregates the outputs of the streaming stage along the time dimension to build a fixed-size embedding of the audio signal.

\subsection{Model Architecture}
\label{subsec:archi}

The sequence-wise network of our solution is pictured in blue on the leftmost part of Figure~\ref{fig:network_diagram}. 
One basic component, taken from the QuartzNet~\cite{quartznet} model, is a Time Channel Separable 1-dimensional Convolution (TCSConv1d) module. It is composed of a 1-dimensional depthwise convolution, where each kernel operates only across the time axis, and a pointwise convolution acting on all filters but independently on each frame.
The first TCSConv1d module is followed by Batch Normalisation (BN) and Parametric ReLU (PReLU)~\cite{prelu}. The next layer performs max-pooling to halve subsequent computations.
Successive blocks are also composed of TCSConv1d followed by BN and a Max Features Map (MFM) operation~\cite{mfm} that, at each location, optimally selects the output of distinct filters. As shown in Figure~\ref{fig:network_diagram}, the TCSConv1d+BN followed by MFM constitutes a block that is repeated $R$ times with a residual connection and followed by a PReLU activation.

Time aggregation techniques are crucial for creating efficient embeddings. Historically, aggregation was performed using a simple mean pooling mechanism~\cite{tap}, later refined by statistical pooling~\cite{sp} and attention mechanisms~\cite{asp, sap}, or using that last output of the recurrent cells~\cite{heigold_end--end_2015}. Statistical pooling being too demanding for our constrained budget, we compared Self Attentive Pooling (SAP), recurrent cells, and GLVAD and got better results with the latter.
To further limit the computations, we replace the last linear projection in GVLAD by a simple cluster-wise projection averaged on the cluster dimension $K$. This comes with almost no performance drop, while making the cost of adding a new cluster linear instead of quadratic in the number of learning weights in the projection layer. 
The rightmost part of Figure~\ref{fig:network_diagram} (in yellow) displays a detailed view of the embedding-wise network. $O=K+G$ is the total number of clusters and is defined as the sum of the numbers of clusters $K$ contributing to the embedding and the number of "Ghost" clusters $G$, named that way because they are not included in the final concatenation (see~\cite{gvlad} for more details). 

\subsection{Training loss function}
\label{subsec:loss}

Speaker recognition can be seen as a metric learning or a classification problem; both approaches have been shown to be successful. 
We explored triplet loss techniques for metric learning (like soft triplet loss based on softmax~\cite{triplet_soft}), and cross-entropy based methods for classification, like 
angular prototypical~\cite{angular_prototypical} or ArcFace~\cite{arcface} losses. 
We found that ArcFace was the most efficient, especially when coupled with focal loss~\cite{focal_loss}. 

\section{Experiments}
\label{sec:experiments}

\subsection{Datasets}
\label{subsec: dataset}

The proposed approach is evaluated on two datasets. We first train our network on the VoxCeleb2~\cite{voxceleb2} dev split only. This dataset contains $5994$ speakers of $145$ different nationalities, with over $1,092,009$ utterances.
We compared this model with previously published papers in Section~\ref{subsec:results_clean} using the VoxCeleb1 verification test, following the protocol proposed by~\cite{voxceleb2}.

We also train an additional model using the same architecture, but augmenting the training data with the MUSAN~\cite{musan} dataset, a corpus of music, speech, and noise.
Reverberation effects are obtained by applying rooms simulation from Pyroomacoustics~\cite{Scheibler2017Oct}. We simulate over $5,000$ rooms and apply the augmentation randomly eight times for each original audio sample with variation of gain, SNR, noise type, and noise source location. We also include a small amount of speed augmentation as a approximate means of accounting for within-speaker speech tempo variability.
The model trained with augmented data is analysed 
in challenging acoustic conditions in Section~\ref{subsec:results_noisy} on the Voices Obscured in Complex Environmental Settings (VOiCES) corpus~\cite{richey2018voices}. The corpus addresses challenging noisy and far-field acoustic environments known to strongly impact the final performance of speaker verification systems.

\subsection{Experimental setup}

\subsubsection{Model implementation}
The acoustic features are $64$-dimensional Mel filterbank energies, extracted from the input audio every $10$ms over a window of 20ms. Mean and variance normalisation (MVN) is applied but no Voice Activity Detection (VAD) pre-processing is done. We experimented with several combinations of depth and width of the architecture (while keeping the total number of parameters fixed) and converged to a configuration with a total of $22$ TCSConv1d each with $96$ filters ($B=5$ and $R=3$ in Figure~\ref{fig:network_diagram}). All the kernels have a constant size of $15$, tuned to reach a good balance between performance, computational load and final size of the receptive field.

Compared with the original GVLAD work~\cite{gvlad}, we select a higher number of clusters $K$ at the aggregation stage (see Section~\ref{subsec:archi}) as we find it yields better results. The number of Ghost clusters $G$ seems to have little impact as soon as there are at least 3. We therefore set $K=32$ and $G=3$ for the following experiments. The embedding network has $237,499$ learning parameters and an output dimension of $96$. Despite our constrained budget, we do not apply compression techniques such as quantization, teacher-student, or pruning. We of course expect these refinements to further improve our model, but we choose to focus solely on optimizing the architecture in this paper.

\subsubsection{Training details}

Each batch contains $S$ speakers, each with $N$ utterances of duration $D$ in seconds. Best results are obtained with: $S=75$, $N=5$, and $D=2\mathrm{to}5$. $2\mathrm{to}5$ means a random uniform sampling of each sample from 2 to 5 seconds. 
The same definition of training epoch as in~\cite{angular_prototypical} is used: every speaker in the dataset is seen once. Given the memory constraints, we train with mixed precision and auto scaling loss which we find useful to stabilise the training and avoid exploding gradients. 
We use the Adam optimizer with a learning rate of $0.001$ and a weight decay of $0.0005$. A scheduler is applied to reduce the learning rate on loss plateau while monitoring the EER on the VoxCeleb2 test set. The scheduler patience is set to $250$ epochs.
The scaling and margin parameters of the ArcFace~\cite{arcface} loss function are respectively set to $s=15$ and $m=0.5$. 
Masking on the time-dimension is applied on the sequence-wise part and during the first stages of the GVLAD process.

\subsection{Inference \& Streaming}

\begin{table}
    \centering
     \begin{tabular}{c| c c c c} 
     \toprule
     Network Part & FMA* & Div* & FLOPS & params \\ [0.5ex] 
     \midrule
     
     Sequence wise & $10850.4$ & $0$ & $10.8$M & $211.6$K  \\ 
     Embedding wise & $659.7$ & $8.0$ & $0.7$M &  $25.8$K \\
     \midrule
     TOTAL & $11509.4$ & $8.0$ & $11.5$M &  $237.5$K \\
     \bottomrule
    \end{tabular}
    \caption{Computational Inference Cost (* K over 1s).
    \textit{FMA}: fused multiply–add operation.
    \textit{Div}: div operation. 
    \textit{FLOPS}: floating point operations per second.
    \textit{params}: number of learning parameters.
    }
    \label{fig:computation_cost_table}
\end{table}

Separating the model in two distinct stages allows to set a decoupled memory and latency budget. Thanks to \texttt{tract}\footnote{available at \url{ https://github.com/sonos/tract} version used: \texttt{0.11.1}}, an open source neural network inference library, we are able to run the first stage in streaming mode and drastically reduce the delay after end-pointing. Some numbers for each stage of the proposed approach are displayed in Table~\ref{fig:computation_cost_table}. 

Latency is a cornerstone to real-time applications of speaker verification systems. It depends on the audio duration to embed and the computational power given a fixed model. Both dimensions are displayed in Table~\ref{fig:latency}. These results show that the system can run on a single core of Raspberry Pi3 B. This latency could be further improved by streaming part of the embedding-wise network
(dashed orange box in the rightmost part of Figure~\ref{fig:network_diagram}). This will be the object of a future work.

\begin{table}
    \centering
     \begin{tabular}[width=\textwidth]{c| c c c c c} 
     \toprule
      CPU \& inference & 1s & 5s & 10s \\ [0.5ex] 
     \midrule
     Intel i7-8750H - Batch & 14.8 &  33.7 & 58.40 \\ 
     Intel i7-8750H - Stream & 1.11 & 5.09 & 9.97 \\
     \midrule
     Raspberry Pi 3B - Batch & 198.8 & 438.2 & 733.8 \\
     Raspberry Pi 3B - Stream & 45.1 & 184.4 & 221.5 \\
     \bottomrule
    \end{tabular}
    \caption{Mean embedding latency in milliseconds on a mono-core CPU (Raspberry Pi running on 64-bit Ubuntu) for various audio lengths,
    \texttt{Batch} is the latency if we apply the whole embedding process once end-pointing is triggered,
    \texttt{Stream} is the latency when the first stage is performed in streaming with a Real Time Factor (RTF) lower than 1.
    }
    \label{fig:latency}
\end{table}

\section{Results}
\label{sec:results}

The proposed approach is evaluated by computing a cosine similarity score between embeddings on two datasets.

\subsection{VoxCeleb1 verification test}
\label{subsec:results_clean}

\begin{figure}[ht!]
    \centering
    \includegraphics[scale=0.65]{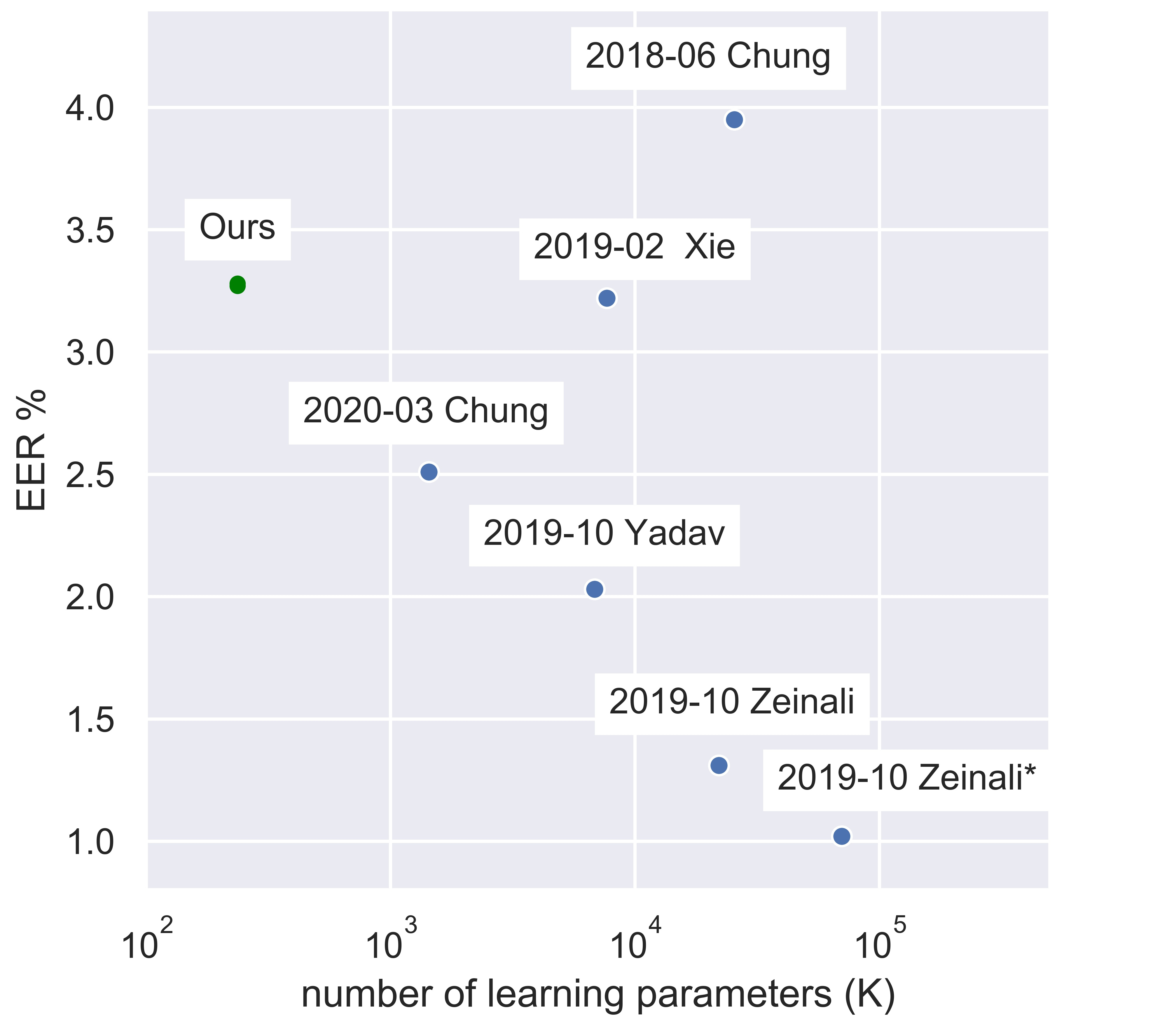}
    \caption{
        EER\% comparison on the VoxCeleb1 verification test 
        with several previous approaches:
        \texttt{2018-06 Chung}~\cite{voxceleb2}, \texttt{2019-02 Xie}~\cite{gvlad}, \texttt{2020-03 Chung}~\cite{angular_prototypical}, 
        \texttt{2019-10 Yadav}~\cite{Yadav2019Oct}, \texttt{2019-10 Zeinali}~\cite{Zeinali2019Oct}
        , \texttt{2019-10 Zeinali*}~\cite{Zeinali2019Oct} (models fusion). EER for \texttt{2020-03 Chung} has been recomputed by building an embedding model from the provided source code, adding a cosine distance scorer and running the benchmark, all others EERs are reported from the corresponding papers.
    }
    \label{fig:research_comparison}
\end{figure}

The proposed model (trained on the VoxCeleb2 dev split) is compared to other existing works in terms of EER and number of parameters, following the VoxCeleb1 verification test protocol proposed in~\cite{voxceleb2}. Figure~\ref{fig:research_comparison} shows that despite its constrained budget, our model is almost on par with the original GVLAD approach~\cite{gvlad} with 32 times less number of parameters. Other solutions have lower EER, but require computational capabilities incompatible with embedded systems. 
Contrary to the number of parameters, the computation cost is more rarely reported in the literature, but critical to real life applications. The second smallest model displayed on Figure~\ref{fig:research_comparison}, denoted \texttt{2020-03 Chung}, has only $1.437$M learning parameters ($6$ times more than the proposed approach) but actually requires $353.3$ MFLOPS, or $30$ times more operations at processing time than our solution\footnote{To compute these numbers, we built the ONNX model from the source code available at \url{https://github.com/clovaai/voxceleb_trainer}}.

\subsection{VOiCES 2019 challenge}
\label{subsec:results_noisy}

We compare the proposed approach with existing works following the \textit{fixed} training conditions of the VOiCES benchmark (see \cite{Nandwana2019Feb} for more details). The results reported in Table~\ref{table:voices_performance_table} for our approach refer to a  model trained on VoxCeleb2 augmented as detailed in Section~\ref{subsec: dataset}. Even in challenging acoustic conditions, the increase in EER compared to other approaches remains limited ($+2.57$ percentage points from the best performing approach) while the number of parameters is drastically reduced ($26$ times smaller). It should be noted that the best scoring systems in Table~\ref{table:voices_performance_table} employ additional VAD and scoring mechanisms that are expected to significantly improve accuracy especially in challenging conditions. 

\begin{table}
    \centering
    \begin{tabular}[width=\textwidth]{c| c c } 
    \toprule
    Model &  EER\% & params (K) \\ 
    \midrule
    STC-Innovations Ltd. \cite{Novoselov2019Apr} & 5.04 & 5953.5\\
    BUT from Brno University \cite{Zeinali2019Jul} & \textbf{4.90} & 6083.0 \\

    Ours & 7.47 & \textbf{237.5}\\
    \bottomrule
    \end{tabular}
    \caption{Comparison in the \textit{fixed} training conditions~\cite{Nandwana2019Feb} of EER\% and learning parameters to the best reported single models on the VOiCES from a Distance 2019 Challenge.}
    \label{table:voices_performance_table}
\end{table}

\section{Conclusion}
\label{sec:conclusion}
We propose an efficient model architecture for speaker verification suited for embedded systems. Our results demonstrate that our solution yields a limited increase of EER on well established benchmarks, while drastically reducing the number of parameters and operations. Rarely reported in the literature, the inference properties of the model have been studied, highlighting a good level of responsiveness. Future work will be centered on further improving the accuracy in challenging acoustic environments, first by integrating the proposed solution with a low-resource VAD and advanced post-embedding scoring techniques as in~\cite{Zeinali2019Jul}.

\bibliographystyle{IEEEbib}
\bibliography{refs}

\end{document}